\documentclass[aps,prl,reprint,superscriptaddress,nofootinbib]{revtex4-2}
\usepackage[T1]{fontenc}
\usepackage[utf8]{inputenc}
\usepackage{graphicx}
\usepackage{amsmath,amssymb,bm}
\usepackage{hyperref}
\hypersetup{colorlinks=true,citecolor=blue,linkcolor=blue,urlcolor=blue}

\newcommand{\kT}{k_\perp}
\newcommand{\qT}{q_T}
\newcommand{\QM}{Q_M}
\newcommand{\Rshape}{R_{\rm Cu/p}^{\rm shape}}

\begin{document}

\title{Evidence for \texorpdfstring{$Q$}{Q}-Dependent Nuclear Transverse-Momentum Redistribution Beyond Broadening from AI-driven analysis of \texorpdfstring{$p$}{p}--Cu Drell--Yan}

\author{I.~P.~Fernando}
\email{ishara@virginia.edu}
\affiliation{Department of Physics, University of Virginia, Charlottesville, Virginia 22904, USA}

\author{D.~Keller}
\email{dustin@virginia.edu}
\affiliation{Department of Physics, University of Virginia, Charlottesville, Virginia 22904, USA}

\date{\today}

\begin{abstract}
We extract a target-side Cu transverse-momentum profile from fixed-target $p$--Cu Drell--Yan data by holding a momentum-space proton reference fixed and training only an asymmetric Cu kernel in the small-$\qT$ region. In the supported window, $0.15\le x_{\rm Cu}\le0.46$ and $7.5\le\QM\le15.75$ GeV, the nuclear modification is not a universal width increase. It appears as $Q$-dependent redistribution: an $\mathcal{O}(1~{\rm GeV})$ shoulder and compensating probability flow between shoulder and resolved-tail regions, beyond one-parameter broadening.
\end{abstract}

\maketitle

At small $\qT \ll Q$, Drell--Yan (DY) production probes the transverse structure of the incoming hadrons through transverse-momentum-dependent (TMD) parton distributions~\cite{CollinsSoperSterman1985,Collins2011,EIS2012,ScimemiVladimirov2018}. Recent unpolarized TMD phenomenology has been driven by flexible extractions performed in impact-parameter space~\cite{MAP2024,Moos2025}, while machine-learning methods have begun to enter TMD phenomenology, first for polarized TMDs \cite{FernandoKeller2023}, and more recently for unpolarized TMDs~\cite{FernandoKeller2025,MAPNN2025}. Here we use the momentum-space framework of Ref.~\cite{FernandoKeller2025} to isolate the Cu target-side TMD profile directly from fixed-target proton--Cu data, without introducing hydrogen/deuterium ratios or external global nuclear-fit information into the Cu training.

Nuclear modifications of DY transverse-momentum spectra have a long experimental and phenomenological history. NA10 observed a nuclear dependence of the transverse-momentum distribution of massive dimuon pairs in W/D comparisons~\cite{Bordalo1987}, and Fermilab E772/E866 measurements and reanalyses established DY transverse-momentum broadening as a probe of initial-state multiple scattering in cold nuclear matter~\cite{Johnson2007,Kang2016}. Global nuclear-PDF analyses describe the accompanying shadowing, antishadowing, and EMC effects~\cite{EPPS21,nCTEQ15,KlasenPaukkunen2024}. More recently, global nTMD analyses extracted nuclear-modified TMDs from SIDIS and DY data using broadening-type nonperturbative parameters and scale-evolution extensions~\cite{Alrashed2022,Alrashed2023}, while fixed-target DY studies reported evidence for a transverse EMC effect~\cite{Barry2023}. First-principles $pA$ TMD work further separates possible intrinsic nuclear TMD modifications from perturbative in-medium dynamics~\cite{Ke2025}. The question addressed here is therefore not whether nuclei modify transverse spectra, but whether fixed-target $p$--Cu data require a target-side, momentum-space redistribution of the TMD profile that is more differential than a single broadening parameter.

The proton baseline used here comes from the two-stage extraction of Ref.~\cite{FernandoKeller2025}, which learns the fixed-target DY kernel and reconstructs a normalized profile
\begin{equation}
 f_{1,q/h}(x,\kT;Q,Q^2)=f_{1,q/h}(x;Q^2)\,s_h(x,\kT;Q),
\label{eq:tmdprofile}
\end{equation}
with $\int d^2{\bf k}_\perp\,f_{1,q/h}(x,\kT;Q,Q^2)=f_{1,q/h}(x;Q^2)$ at each $(x,Q)$, where the subscript $h$ represents the parent hadron. The baseline analysis was carried out natively in $\kT$ space with a tree-level hard factor, no explicit $Y$ term, and $Q$ dependence learned directly from small-$\qT$ data~\cite{FernandoKeller2025}. The present Letter holds the resulting proton profile fixed on the beam leg and extracts a per-nucleon Cu profile from proton--copper data dominated by E605~\cite{Moreno1991}.

The Cu isolation follows the same two stages. Stage I learns, replica by replica, a reduced proton--copper {\it transverse structure kernel}  $S_{p{\rm Cu}}(\qT;x_p,x_{\rm Cu},\QM)$ after dividing the measured cross section by the known tree-level prefactor and charge-weighted collinear PDF combinations. Stage II determines a normalized target-side Cu profile ({\it intrinsic- transverse-momentum profile}) from the asymmetric convolution
\begin{equation}
\begin{split}
S_{p{\rm Cu}}(\qT;x_p,x_{\rm Cu},Q)=\int_0^\infty\!dk\,k\int_0^{2\pi}\!d\phi\;s_p(x_p,k;Q)\\
\times s_{\rm Cu}\!\left(x_{\rm Cu},\sqrt{\qT^2+k^2-2\qT k\cos\phi};Q\right).
\end{split}
\label{eq:pcuconv}
\end{equation}
For each matched replica, $s_p^{(r)}$ is held fixed while only $s_{\rm Cu}^{(r)}$ is trained to reproduce the reduced $S_{p{\rm Cu}}$ kernel across the measured $(x_p,x_{\rm Cu},\QM,\qT)$ bins. No beam-exchange symmetrization is imposed: the inverse problem is an asymmetric deconvolution rather than an auto-convolution. The Stage-I reduction samples NNPDF4.0 proton-PDF replicas~\cite{Ball2022} through LHAPDF~\cite{Buckley2015LHAPDF} on both legs, so $s_{\rm Cu}$ is the effective target-side transverse deformation required by the Cu data relative to a fixed proton reference, rather than a fitted nuclear-PDF parametrization.

For E605 kinematics, the published $(\QM,x_F,\qT)$ bins are converted at leading power to $(\QM,y,\qT)$ and then to $(x_p,x_{\rm Cu})$ using $x_{p,{\rm Cu}}=(Q/\sqrt{s})e^{\pm y}$. The same small-$\qT$ cut, scale choice $\mu=\sqrt{\zeta}=Q$, taper normalization, and replica protocol are retained from the proton fit, and all Cu/p observables are formed replica by replica from matched proton/Cu ensembles. Since the Cu data do not constrain the full $(x,Q)$ plane uniformly, all quantitative claims are restricted to the fiducial window, excluding the $\Upsilon$ region,
\begin{equation}
0.15\le x_{\rm Cu}\le0.46,\qquad 7.5\le \QM\le15.75~{\rm GeV},
\label{eq:fiducial}
\end{equation}
which is the kinematic support of the Cu training data.

\begin{figure*}[t]
  \centering
  \includegraphics[width=0.96\textwidth]{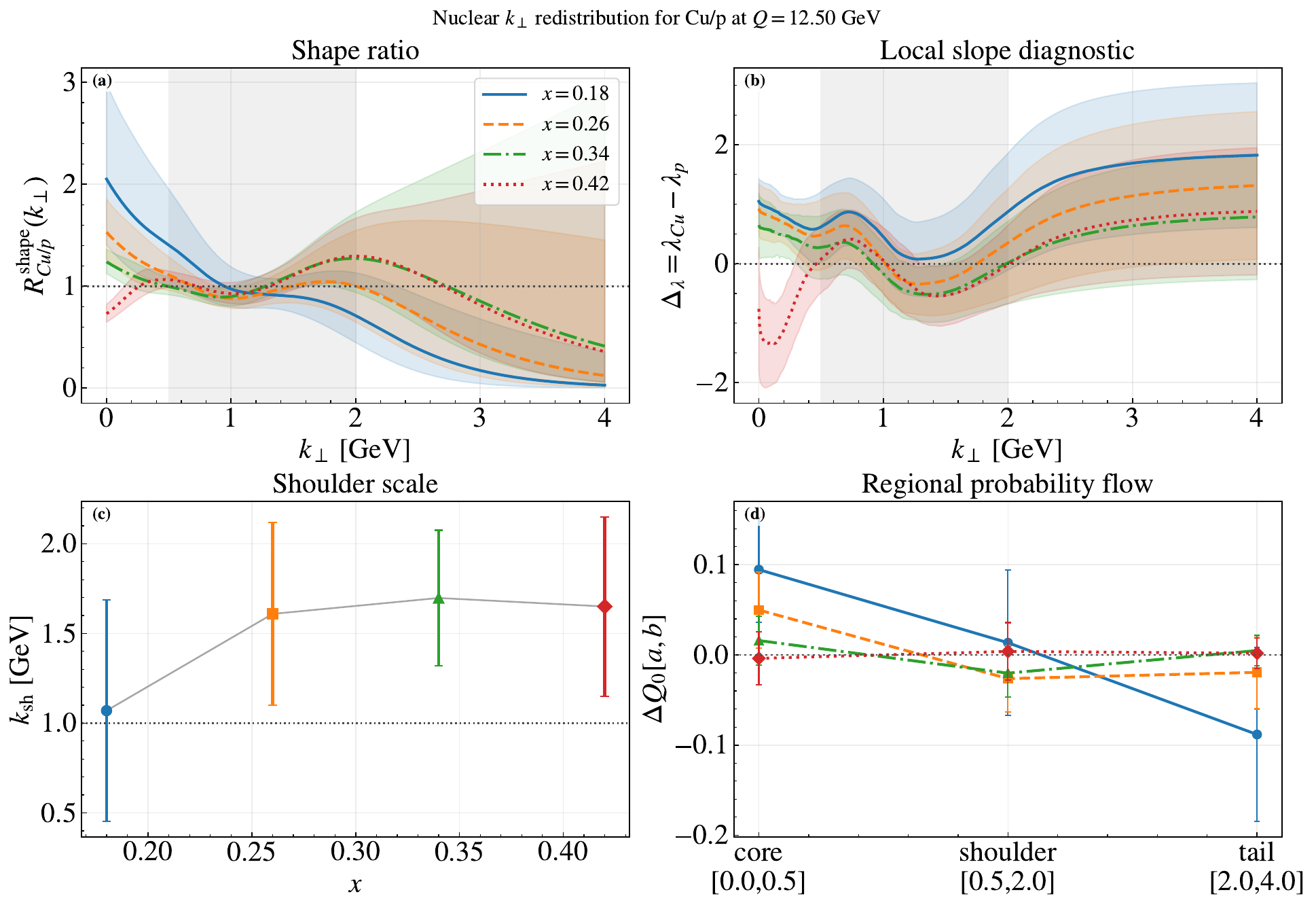}
  \caption{Representative fixed-$Q$ anatomy of the Cu/p nuclear modification within the fiducial region. Panel (a) shows the shape ratio $\Rshape$, panel (b) the local-slope diagnostic $\Delta_\lambda$, panel (c) the extracted shoulder scale $k_{\rm sh}$, and panel (d) the regional probability flow $\Delta Q_0[a,b]$ in the core, shoulder, and resolved-tail windows. The local enhancement in $\Rshape$ is concentrated in an intermediate-$\kT$ shoulder, while the integrated compensation is distributed differently across the three windows.}
  \label{fig:hero}
\end{figure*}

\begin{figure*}[t]
  \centering
  \includegraphics[width=0.99\textwidth]{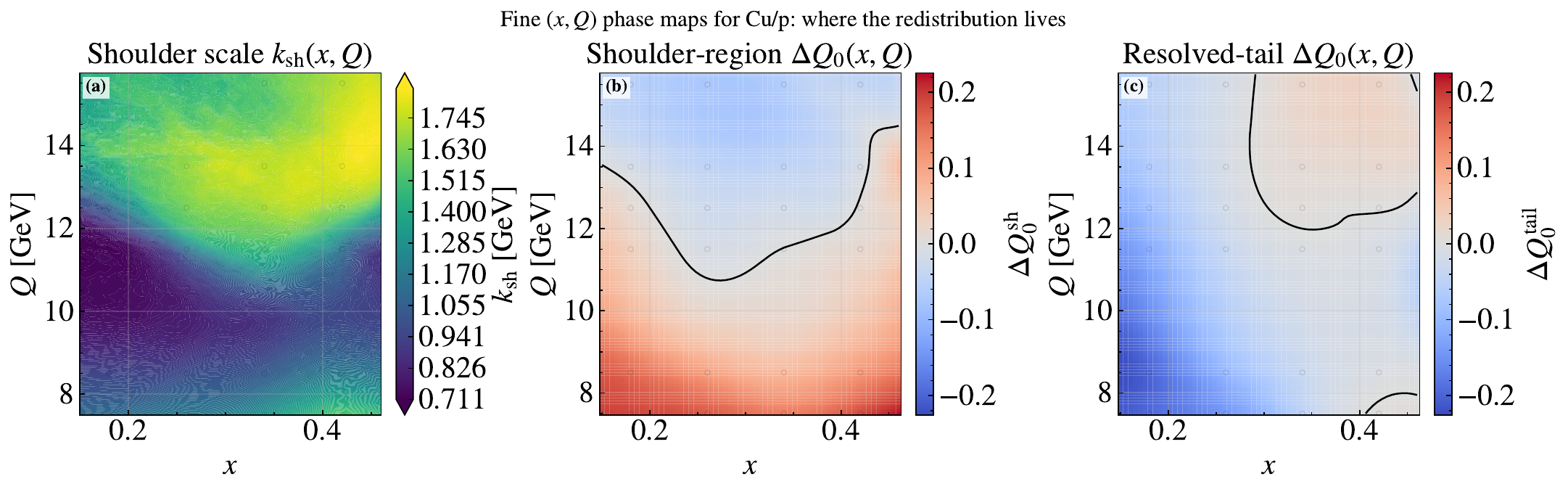}
  \caption{$(x,Q)$ phase maps for the Cu/p redistribution over the Cu-supported region of Eq.~(\ref{eq:fiducial}). Panel (a) shows the shoulder scale $k_{\rm sh}(x,Q)$, panel (b) the shoulder-region flow $\Delta Q_0^{\rm sh}$, and panel (c) the resolved-tail flow $\Delta Q_0^{\rm tail}$. Both the location of the local enhancement and the balance of regional probability flow evolve across the measured mass window.}
  \label{fig:phase}
\end{figure*}

To isolate shape rather than normalization, we compare proton and Cu profiles through a common numerical shape normalization $\tilde s_h$. For the comparisons below, the two learned profiles are evaluated at a common momentum fraction denoted by $x$, and we form
\begin{equation}
 \Rshape(x,\kT;Q)=\frac{\tilde s_{\rm Cu}(x,\kT;Q)}{\tilde s_p(x,\kT;Q)}.
\label{eq:rshape}
\end{equation}
A one-parameter broadening scenario would produce a largely monotonic modification and width-like observables with a fixed sign. The Cu extraction instead shows a localized shoulder in the intermediate region $0.5\lesssim\kT\lesssim2$ GeV, as seen in Fig.~\ref{fig:hero}. We quantify its location by
\begin{equation}
 k_{\rm sh}(x,Q)=\arg\max_{0.5\le\kT\le2.0~{\rm GeV}}\Rshape(x,\kT;Q),
\label{eq:ksh}
\end{equation}
and define the regional probability-flow integral
\begin{equation}
\begin{split}
 \Delta Q_0[a,b](x,Q) = {} & 2\pi\!\int_a^b\!d\kT\,\kT \\
 & \times \bigl[\tilde s_{\rm Cu}(x,\kT;Q)-\tilde s_p(x,\kT;Q)\bigr],
\end{split}
\label{eq:dq0}
\end{equation}
which measures whether a given $\kT$ window gains or loses probability relative to the proton. We evaluate Eq.~(\ref{eq:dq0}) in the core $[0,0.5]$ GeV, shoulder $[0.5,2.0]$ GeV, and resolved tail $[2.0,4.0]$ GeV. Error bars in Fig.~\ref{fig:hero} and smooth bands in Fig.~\ref{fig:phase} are 68\% central intervals from 1000 matched replicas.

The fixed-$Q$ anatomy in Fig.~\ref{fig:hero} shows that the shape ratio develops a local shoulder rather than a rigid shift. The local-slope diagnostic, $\lambda_h(x,\kT;Q)\equiv-\partial_{\kT}\ln\tilde s_h$ with $\Delta_\lambda=\lambda_{\rm Cu}-\lambda_p$, changes sign and magnitude across $\kT$, rather than acting as a constant width renormalization. The extracted $k_{\rm sh}$ values are of order 1 GeV, while the regional-flow panel shows why a single full-range width moment is incomplete: a pronounced local enhancement in $\Rshape$ can coexist with shoulder-window flow that is positive, nearly zero, or negative, depending on the kinematics.

The redistribution is also $Q$ dependent. Fig. \ref{fig:phase} summarizes the supported $(x,Q)$ plane. The shoulder scale $k_{\rm sh}(x,Q)$ drifts across the measured mass window, so the nuclear modification is not tied to one fixed transverse scale. At the same time, the shoulder-region flow and resolved-tail flow vary differently with $Q$ and may change sign across the fiducial domain. The Cu modification therefore cannot be reduced to a scale-independent broadening parameter extracted from a single moment. This behavior is sharper than in the proton baseline, whose fixed-target $Q$ variation was mild over the same lever arm~\cite{FernandoKeller2025}.

The interpretation can also be stated in a model-independent way. At leading power, the unpolarized small-$\qT$ proton--nucleus cross section can be written schematically as
\begin{equation}
\begin{split}
\frac{d\sigma^{pA}}{dQ\,dy\,d^2{\bf q}_T}\propto
\sum_q C_q(Q)\bigl[&S(\qT,x_a,x_b;Q)\\
&+S(\qT,x_b,x_a;Q)\bigr],
\end{split}
\label{eq:exchange}
\end{equation}
with $C_q(Q)=e_q^2$, $x_{a,b}=(Q/\sqrt{s})e^{\pm y}$, and the unsymmetrized kernel $S$ built from one beam-side and one target-side TMD~\cite{Collins2011,EIS2012}. The full unpolarized cross section is beam-exchange even, but the unsymmetrized kernel need not be in $pA$. Forward/backward differences probe the exchange-odd combination and thus provide a future test of the redistribution pattern in rapidity-asymmetric proton--nucleus observables~\cite{PaukkunnenSalgado2011,CMSpPbDY2021,ATLASpPbZ2015,LHCbpPbZ2023}. The $Q$ dependence seen here is therefore not a trivial common evolution factor; it reflects a difference between the learned proton and nuclear transverse sectors.

The present Letter is intentionally narrow: it is not a full global, resummed TMD analysis, but a data-driven statement about the small-$\qT$ fixed-target regime obtained in the same momentum-space framework that established the proton baseline~\cite{FernandoKeller2025}. The novelty is not the existence of nuclear broadening, which earlier DY measurements already established, nor the first extraction of nuclear-modified TMDs, which global nTMD analyses have performed. Rather, within the Cu-supported domain, the fixed-target $p$--Cu data resolve a target-side, $Q$-dependent transverse-momentum redistribution: an intermediate-$\kT$ shoulder accompanied by compensating regional probability flow. This is a more differential nuclear signature than a single width or broadening parameter.

\begin{acknowledgments}
The authors acknowledge Research Computing at the University of Virginia for providing computational resources and technical support. This work was supported by the U.S. Department of Energy under contract DE-FG02-96ER40950.
\end{acknowledgments}

\end{document}